\documentclass[3p,times,twocolumn,fleqn]{elsarticle}

\biboptions{comma,sort&compress}
\usepackage{graphicx}
\usepackage{here}
\usepackage{ecrc}
\volume{00}
\firstpage{1}
\journalname{Nuclear and Particle Physics Proceedings}
\runauth{}
\jid{nppp}
\jnltitlelogo{Nuclear and Particle Physics Proceedings}
\usepackage{amssymb}
\usepackage[figuresright]{rotating}

\usepackage{amsmath}
\usepackage{braket}
\usepackage{placeins}

\newcommand{\beq}{\begin{eqnarray}}
\newcommand{\eeq}{\end{eqnarray}}
\newcommand{\beqnn}{\begin{eqnarray*}}
\newcommand{\eeqnn}{\end{eqnarray*}}
\newcommand{\Tr}{\ensuremath{\mathrm{Tr}}}

\newcommand{\cool}{\mathrm{cool}}
\newcommand{\sphal}{\mathrm{Sphal}}
\newcommand{\Cov}{\mathrm{Cov}}

\begin{document}

\begin{frontmatter}
	
\title{Sphaleron rate from lattice QCD $^*$}
	
\cortext[cor0]{Talk given at 26th International Conference in Quantum Chromodynamics (QCD 23),  10 - 14 july 2023, Montpellier - FR}
	
\author[aff1]{Claudio Bonanno\fnref{fn1}}
\fntext[fn1]{Speaker, Corresponding author.}
\ead{claudio.bonanno@csic.es}

\author[aff2]{Francesco D'Angelo}
\ead{francesco.dangelo@phd.unipi.it}
\author[aff2]{Massimo D'Elia}
\ead{massimo.delia@unipi.it}
\author[aff2]{Lorenzo Maio}
\ead{lorenzo.maio@phd.unipi.it}
\author[aff2]{Manuel Naviglio}
\ead{manuel.naviglio@phd.unipi.it}

\address[aff1]{Instituto de F\'isica Te\'orica UAM-CSIC, c/ Nicol\'as Cabrera 13-15, Universidad Aut\'onoma de Madrid, Cantoblanco, E-28049 Madrid, Spain}
\address[aff2]{Dipartimento di Fisica dell'Universit\`a di Pisa \& INFN Sezione di Pisa, Largo Pontecorvo 3, I-56127 Pisa, Italy}

\pagestyle{myheadings}
\markright{ }

\begin{abstract}
We compute the sphaleron rate on the lattice from the inversion of the Euclidean time
correlators of the topological charge density, performing also controlled continuum and
zero-smoothing extrapolations. The correlator inversion is performed by means of a
recently-proposed modification of the Backus-Gilbert method.
\end{abstract}


\end{frontmatter}

\section{Introduction}\label{sec:intro}

The QCD sphaleron rate, i.e., the rate of real-time topological transitions due to strong interactions,
\beq\label{eq:rate_def}
\Gamma_\sphal \equiv \underset{t_{\mathrm{M}}\to\infty}{\underset{V_s\to\infty}{\lim}} \, \frac{1}{V_s t_{\mathrm{M}}}\left\langle\left[\int_0^{t_{\mathrm{M}}} d t_{\mathrm{M}}' \int_{V_s} d^3x \, q(t_{\mathrm{M}}', \vec{x})\right]^2\right\rangle
\eeq
with $t_{\mathrm{M}}$ the real Minkowski time and $q(x) = \frac{1}{16\pi^2}\Tr\left\{G_{\mu\nu}(x) \widetilde{G}^{\mu\nu}(x)\right\}$ the QCD topological charge density, is a phenomenologically-relevant quantity, as it has been shown to play a fundamental role in describing the Chiral Magnetic Effect in the quark-gluon plasma~\cite{Fukushima:2008xe, Kharzeev:2013ffa, Astrakhantsev:2019zkr, Almirante:2023wmt}, as well as in describing the axion thermal rate production in the early Universe~\cite{Berghaus:2020ekh,Notari:2022ffe}.

Lattice QCD numerical calculations offer a natural framework to compute $\Gamma_\sphal$ from first principles, and in recent times a few lattice determinations have appeared in the literature~\cite{Kotov:2018aaa, Kotov:2019bt, Altenkort:2020axj, BarrosoMancha:2022mbj}, all limited to the quenched theory due to several non-trivial difficulties one has to face to compute the rate (some of them will be discussed below). This proceedings reports on the main results of~\cite{Bonanno:2023ljc}, where a new strategy to determine $\Gamma_\sphal$ from lattice QCD is proposed. Since the aim of the present paper is to discuss such method and all the systematics involved, the numerical calculations that will be shown in the following will be limited to one temperature and to the pure-gauge theory, where our results can be compared with previous determinations. Instead, the first full QCD determinations of the sphaleron rate, achieved with the methods of~\cite{Bonanno:2023ljc}, can be found in~\cite{Bonanno:2023thi}.

Real-time objects like~\eqref{eq:rate_def} are not amenable to be computed within numerical lattice calculations, which are based on the Euclidean formulation of the theory. Instead, one has to rely on the representation of the sphaleron rate as the zero-frequency limit of the slope of the spectral density of the Euclidean topological charge density correlator $G(t)$:
\beq\label{eq:kubo}
\Gamma_\sphal = 2T \lim_{\omega \to 0} \frac{\rho(\omega)}{\omega},
\eeq
\beq\label{eq:rho_def}
G(t) = - \int_0^{\infty} \frac{d\omega}{\pi} \rho(\omega) \frac{\cosh\left[\omega t - \omega/(2T)\right]}{\sinh\left[\omega/(2T)\right]},
\eeq
\beq\label{eq:def_tcorr}
G(t) \equiv \int d^3 x \braket{q(t, \vec{x}) q(0, \vec{0})}.
\eeq
The density $\rho(\omega)$ is related to the correlator $G(t)$ via the Kubo formula~\eqref{eq:rho_def}.

On the lattice, one cannot directly determine the spectral density $\rho(\omega)$, but just its integral, i.e., the correlator $G(t)$, using a discretized form of~\eqref{eq:def_tcorr}. Thus, determining $\Gamma_\sphal$ from~\eqref{eq:kubo} requires to numerically invert~\eqref{eq:rho_def}. Such issue falls within a wide class of problems known as \emph{inverse problems}~\cite{Rothkopf:2022fyo,Aarts:2023vsf}. Several strategies to address these kind of problems have been proposed in the literature, which have been shown to yield consistent results among them~\cite{Boito:2022njs,Horak:2021syv,DelDebbio:2021whr,Candido:2023nnb,Altenkort:2020axj,Altenkort:2020fgs,Altenkort:2022yhb,Altenkort:2023oms,Tikhonov:1963aaa,Astrakhantsev:2018oue,Astrakhantsev:2019zkr,BackusGilbert1968:aaa,Brandt:2015aqk,Brandt:2015sxa,Hansen:2019idp,ExtendedTwistedMassCollaborationETMC:2022sta,Frezzotti:2023nun,Evangelista:2023fmt}. In this work we will adopt the so-called \emph{HLT method}, i.e., a recent modification of the Backus--Gilbert method~\cite{BackusGilbert1968:aaa} first introduced in~\cite{Hansen:2019idp}.

Another delicate point posed by the lattice calculation of the sphaleron rate is the determination of the correlator~\eqref{eq:def_tcorr}. In order to identify the correct topological background, one has to smooth the lattice configurations drawn from the Monte Carlo to remove UV fluctuations. However, smoothing modifies the short-distance behavior of $G(t)$ below the \emph{smoothing radius} $r_s\propto\sqrt{\text{amount of smoothing}}$. A possible strategy consists of taking the zero-smoothing-radius limit of the correlator, which has to be taken after the continuum limit~\cite{Kotov:2018aaa,Altenkort:2020axj}, and then perform the inversion of the double-extrapolated correlator. This approach has the drawback of working well only for sufficiently large times, and clearly the knowledge of the correlator only above a certain minimum time separation makes the reconstruction of the spectral density more difficult.

In this work instead we follow a different strategy, namely to perform the inversion of finite-lattice-spacing and finite-smoothing-radius correlators, and postpone the double-limit directly on $\Gamma_\sphal$. This strategy is expected to have few advantages: the inversion is expected to be more stable and less noisy, and the dependence on the smoothing radius is expected to be milder. Indeed, the sphaleron rate is defined as a zero-frequency limit, thus is expected to be insensitive to the UV scale introduced by the finite smoothing radius. In particular, one can expect to find a regime in which the sphaleron rate is fairly independent of the smoothing radius, which is reached when the UV scale set by the smoothing radius is well separated by the IR scale of the topological fluctuations which dominantly contribute to $\Gamma_\sphal$, just like what happens with the topological susceptibility as a function of the gradient flow time.

This paper is organized as follows: in Sec.~\ref{sec:methods} we discuss the determination of the topological charge density correlators from the lattice the HLT method to invert them to obtain the sphaleron rate, in Sec.~\ref{sec:res} we show our results for the sphaleron rate, finally in Sec.~\ref{sec:conclu} we draw our conclusions.

\section{Methods}\label{sec:methods}

\subsection{Determination of the correlators}

In this work we consider the simplest discretization of the pure-gauge action $S_{\mathrm{YM}}=\int d^4x \Tr\{G_{\mu\nu}(x)G_{\mu\nu}(x)\}/(4g^2)$, namely the Wilson plaquette action $S_{\mathrm{W}}[U] = - (\beta/N_c) \sum_{n, \nu \ne \mu}\Re\Tr\left\{ \Pi_{\mu\nu}(n)\right\}$, with $\Pi_{\mu\nu}(n) = U_\mu(n)U_\nu(n+a\hat{\mu})U^\dagger_\mu(n+a\hat{\nu})U^\dagger_\nu(n)$ and the inverse coupling $\beta=2N_c/g^2$.

We determine the correlators on 4 lattices with temporal extents $N_t=12,14,16,20$ and spatial extents $N_s=3 N_t$, tuning $\beta$ according to the lattice spacing determinations $a(\beta)$ of~\cite{Necco:2001xg} in order to keep the spatial size $\ell = aN_s \simeq 1.66$~fm and the temperature $T=1/(aN_t)\simeq 357~\text{MeV}\simeq 1.24 T_c$ constant for each ensemble. The lattice topological charge density is determined computing the simplest clover discretization $q_L(n) = (-1/2^9 \pi^2)\sum_{\mu\nu\rho\sigma=\pm1}^{\pm4}\varepsilon_{\mu\nu\rho\sigma}
\Tr\left\{\Pi_{\mu\nu}(n)\Pi_{\rho\sigma}(n)\right\}$ on smoothened configurations, and the time-correlator is simply given by:
\beq
\frac{G_L(tT)}{T^5} = \frac{N_t^5}{N_s^3} \sum_{\vec{n}_{s,1},\,\vec{n}_{s,2}}\braket{q_L(n_{t,1},\vec{n}_{s,1}) q_L(n_{t,2}, \vec{n}_{s,2})},
\eeq
where $tT = \vert n_{t,1}-n_{t,2}\vert/N_t$ is the physical time separation.

\begin{figure}[!b]
\includegraphics[scale=0.4]{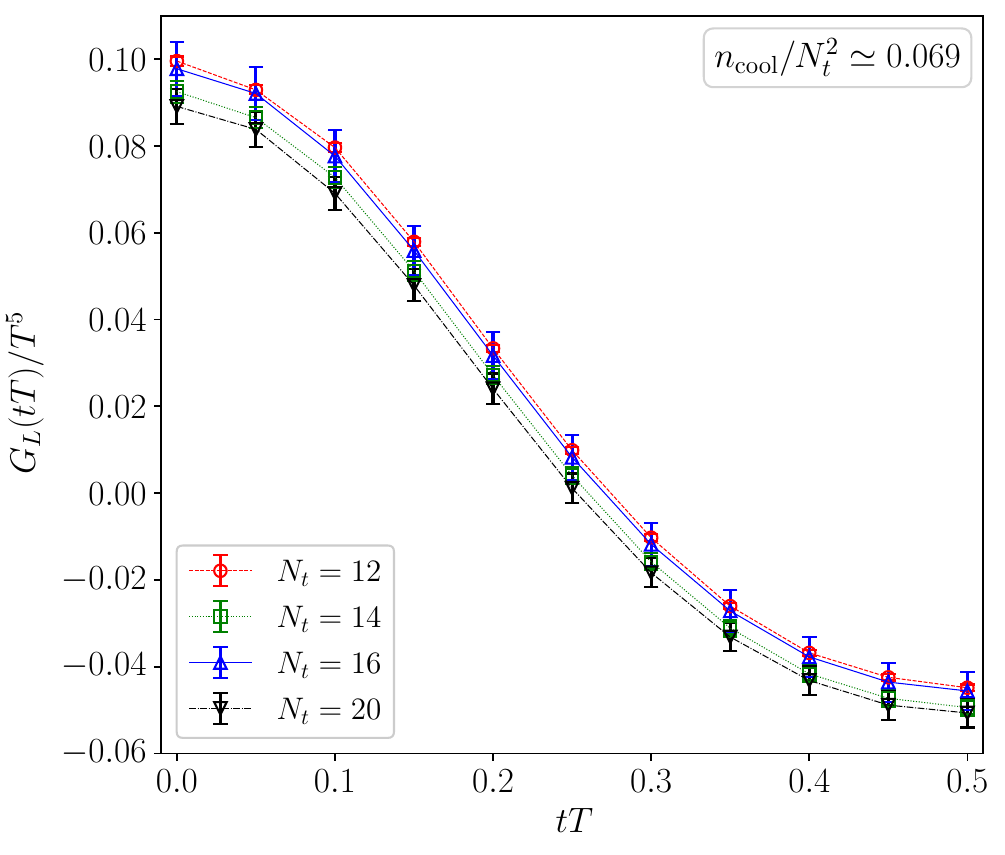}
\caption{Comparison of topological charge density correlators obtained for different lattice spacings, same lattice volume, same temperature and same smoothing radius $(r_s T)^2\propto n_\cool/N_t^2$.}
\label{fig:corrs_comp}
\end{figure}

Examples of obtained lattice topological charge correlators are shown in Fig.~\ref{fig:corrs_comp}. As our smoothing method, we choose cooling~\cite{Berg:1981nw,Iwasaki:1983bv,Itoh:1984pr,Teper:1985rb,Ilgenfritz:1985dz,Campostrini:1989dh,Alles:2000sc} for its simplicity and numerical cheapness. Results obtained with cooling have been shown to be perfectly equivalent to those obtained with other methods~\cite{Alles:2000sc, Bonati:2014tqa, Alexandrou:2015yba}, such as the gradient flow~\cite{Luscher:2009eq, Luscher:2010iy} or stout smearing~\cite{APE:1987ehd, Morningstar:2003gk}. One cooling steps consists in aligning each link to its relative force, so that the Wilson action is locally minimized. The smoothing radius is related to the number of coolings steps $n_\cool$ by~\cite{Bonati:2014tqa}:
\beq
\frac{r_s}{a} = \sqrt{\frac{8}{3}n_\cool},
\eeq
meaning that, with our setup, $\sqrt{n_\cool}/N_t$ is proportional to the smoothing radius expressed in units of $1/T$.

\subsection{The HLT method}\label{sec:HLT}

Let us assume we can write (for unknown $g_t$) this approximation of the spectral density:
\beq\label{eq:Estimator}
\bar{\rho}(\bar{\omega})= - \pi f(\bar{\omega}) \sum_{t=0}^{1/T} g_t(\bar{\omega}) G(t),
\eeq
with $f(\omega) = \omega$, so that
\beq
\frac{\Gamma_\sphal}{2T} = \left[\frac{\bar{\rho}(\bar{\omega})}{\bar{\omega}}\right]_{\bar{\omega}\,=\,0}= - \pi \sum_{t=0}^{1/T} g_t(0) G(t)
\eeq

If we rewrite~\eqref{eq:kubo} as
\beq
G(t) = - \int_0^{\infty} \frac{d\omega}{\pi} \frac{\rho(\omega)}{\omega} K_t'(\omega),
\eeq
where $K'_t(\omega) \equiv \omega \cosh(\omega t - \omega/2T)/\sinh(\omega/2T)$, we obtain the smearing relation
\beq
\frac{\bar{\rho}(\bar{\omega})}{\bar{\omega}} = \int_0^{\infty} d\omega \Delta(\omega,\bar{\omega}) \frac{\rho(\omega)}{\omega},
\eeq
where the resolution function is given by:
\beq
\Delta(\omega,\bar{\omega})= \sum_{t=0}^{1/T} g_t(\bar{\omega}) K'_t(\omega).
\eeq
The more $\Delta(\omega,\bar{\omega})$ is peaked around $\bar{\omega}$ as a function of $\omega$, the better the smeared spectral density $\bar{\rho}(\bar{\omega})$ will approximate the real one $\rho(\omega)$. In the extreme case where the resolution function tends to a Dirac delta: $\Delta(\omega,\bar{\omega}) \to \delta(\omega-\bar{\omega})$ $\implies$ $\bar{\rho}(\bar{\omega})=\rho(\bar{\omega})$. Now we are left with the following problem: how can we determine $g_t(0)$ such that $\Delta(\omega,0)$ is sufficiently peaked around 0 to make $\bar{\rho}(0) \simeq \rho(0)$?

The HLT strategy of~\cite{Hansen:2019idp} to fix the $g_t$ values consists of minimizing the functional:
\beq
F[g_t] = (1-\lambda) A[g_t] +  \frac{\lambda}{\mathcal{C}}B[g_t], \,  \lambda\in [0,1).
\eeq

The term
\beq
A[g_t] = \int_0^{\infty} d\omega \, [\Delta(\omega,0) - \delta_\sigma(\omega,0)]^2 \,e^{2 \omega}
\eeq
depends on the distance between the resolution function and some target function $\delta_\sigma$, which will be discussed in a moment. The term
\beq
B[g_t] = \sum_{t,t'=0}^{1/T} \Cov_{t,t'} \, g_t g_{t'}
\eeq
is instead dependent on the statistical errors on the correlator $G(t)$; finally, $\mathcal{C}$ is just an irrelevant overall coefficient. The regulator parameters $\lambda$ is used to calibrate the contribution of these two terms to the minimized functional $F$. Varying $\lambda$, one can expect two regimes:

\begin{itemize}
\item $\lambda \to 0$: $B[g_t]$ is neglected and the error on $\Gamma_\sphal$ is dominated by large statistical fluctuations, which can be traced back to the fact that the inverse problem we are trying to solve is ill-conditioned.
\item $\lambda \to 1$: $A[g_t]$ is neglected and the error on $\Gamma_\sphal$ is dominated by systematic effects. As a matter of fact, being $\Delta(\omega,0)$ practically unconstrained, we have no control on its shape, thus we cannot guarantee that the obtained $g_t$ coefficients will yield a resolution function sufficiently peaked around $\omega=0$.
\end{itemize}

Thus, we look for an intermediate regime where statistical error dominates over systematic, but a clear signal can be observed for $\Gamma_\sphal$. In particular, we expect to be well into the statistically-dominated regime when $d^2[g_t](\lambda) \equiv (A[g_t]/B[g_t])(\lambda)\ll 1$. Moreover our error bars also keep into account any observed variations of the central values of $\Gamma_\sphal$ within this region.

Finally, let us describe the role of the target function $\delta_\sigma(\omega,0)$. The target function is used to constrain the shape of the resolution function in the minimization process pursued to determine the $g_t$ coefficients. Here we choose:
\beq
\delta_\sigma(\omega,0) = \left(\frac{2}{\sigma \pi}\right)^2 \frac{\omega}{\sinh(\omega/\sigma)},
\eeq
where $\sigma$, the smearing width, is related to the width of the peak of the target function around $\omega=0$. Since $\delta_\sigma(\omega,0)\to\delta(\omega)$ when $\sigma\to 0$, the choice of $\sigma$ affects the quality of our approximation of $\rho$ via $\bar{\rho}$. Smaller values of $\sigma$ will lead to a more peaked resolution function, but the sphaleron rate will be affected by larger errors, as the minimization of the $g_t$ coefficients will be more noisy. On the other hand, larger values of $\sigma$ will allow a more precise determination of the resolution function, which however will lead to smear over a larger region around $\omega=0$, potentially introducing systematic effects. We chose $\sigma/T=1.75$, but verified that no significant difference is obtained choosing other values within the range $1.5 \le \sigma/T \le 2$.

\section{Results}\label{sec:res}

\subsection{Rate from double-extrapolated correlator}\label{sec:res1}

\begin{figure}[!htb]
\includegraphics[scale=0.225]{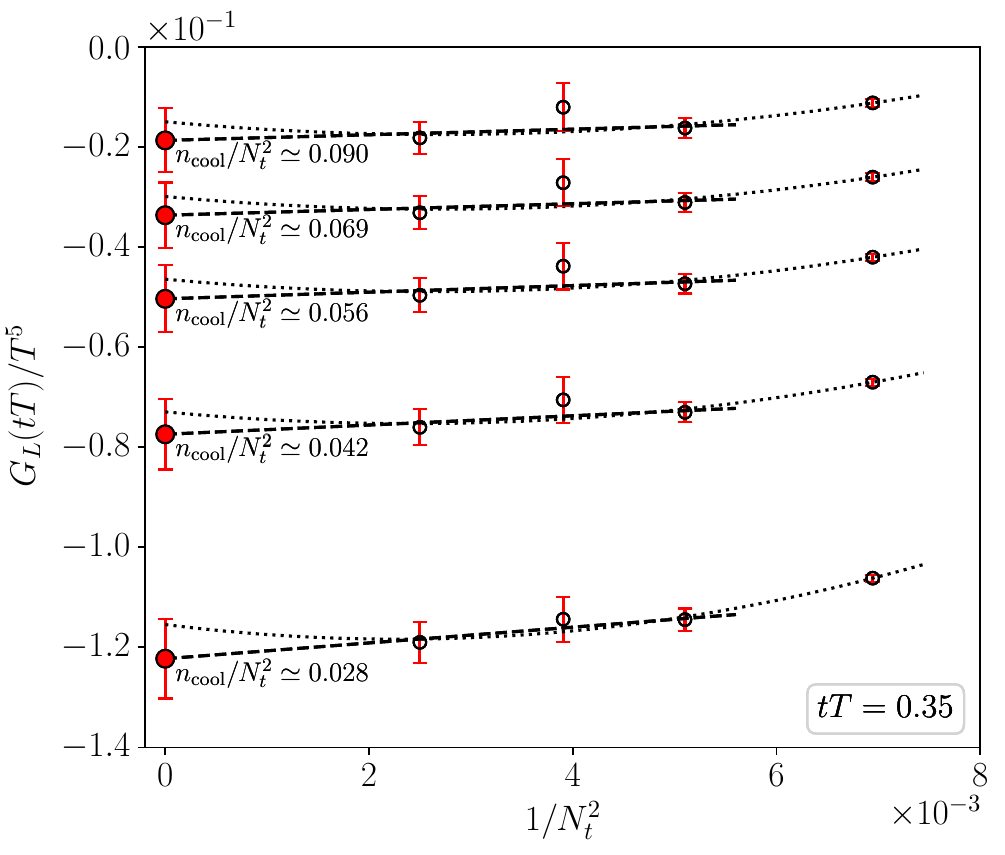}%
\includegraphics[scale=0.225]{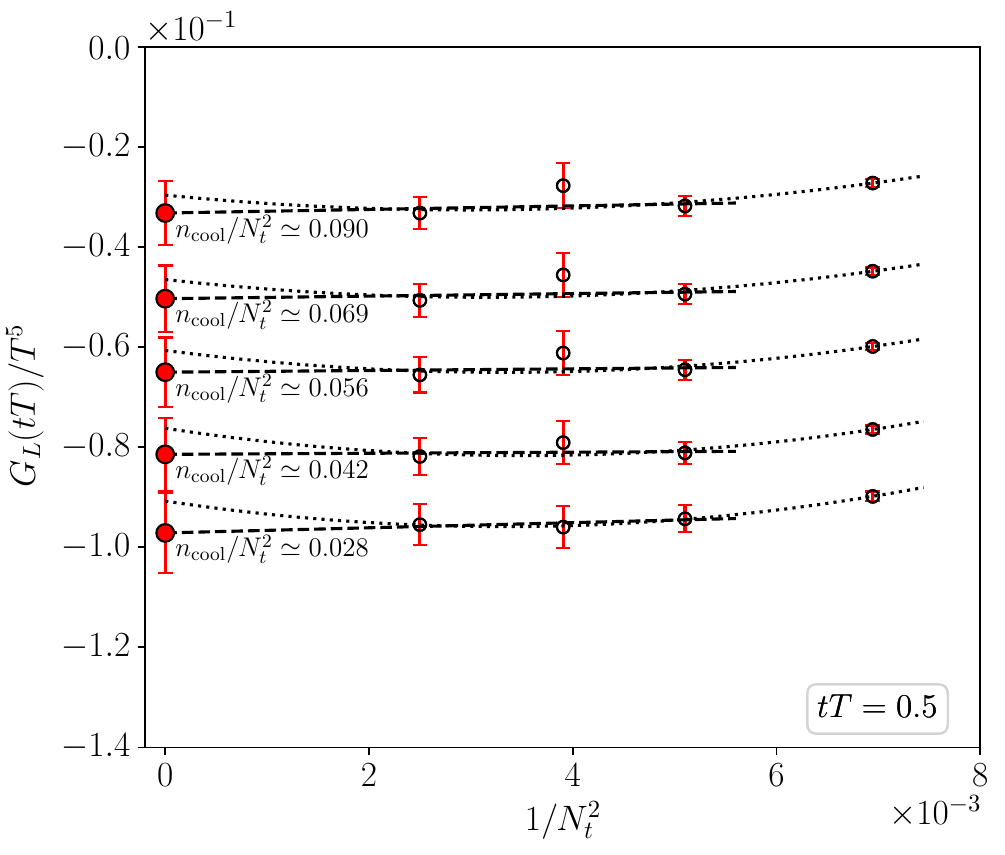}
\includegraphics[scale=0.225]{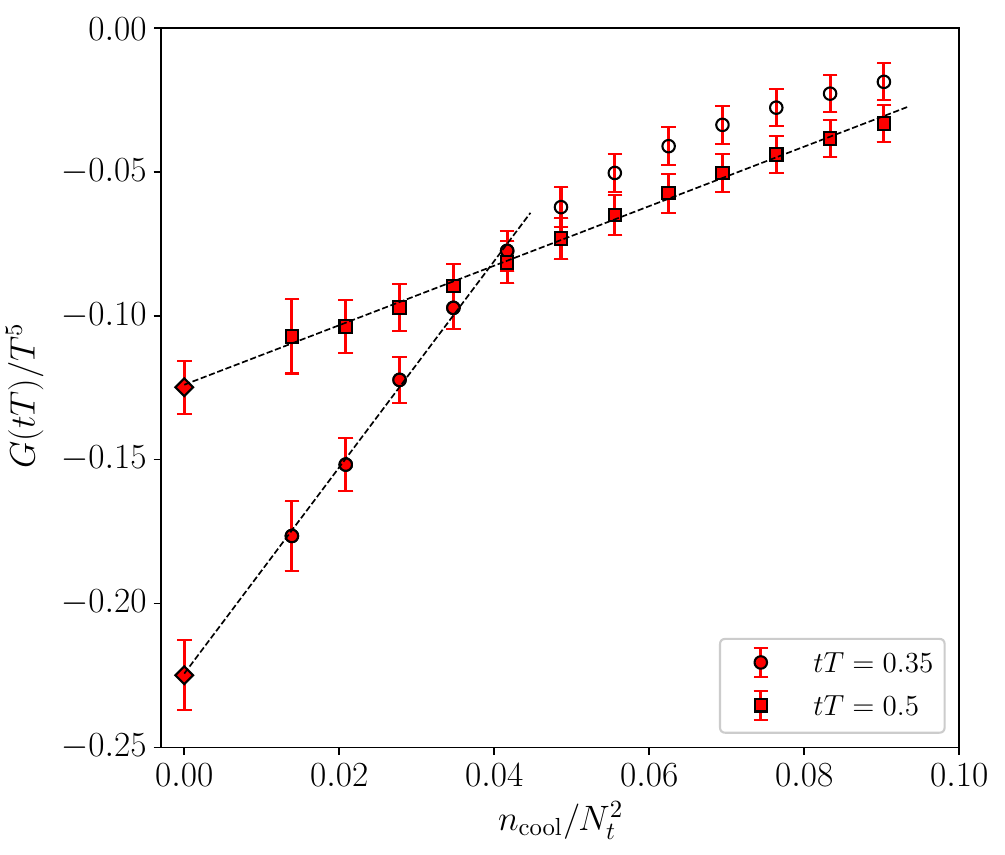}%
\includegraphics[scale=0.225]{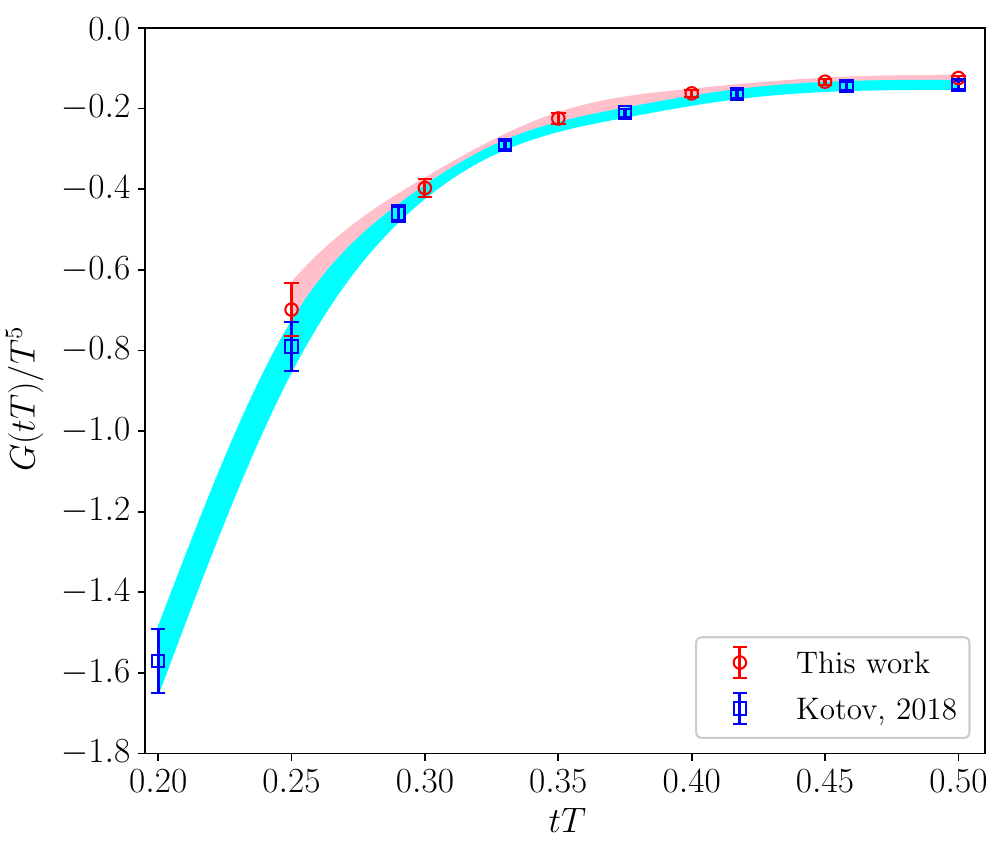}
\caption{Double extrapolation of the topological charge density correlator. Top panels: continuum limit extrapolations at fixed smoothing radius. Bottom left panel: zero-cooling extrapolations. Bottom right panel: double-extrapolated correlator compared with the one obtained in~\cite{Kotov:2018aaa} for the same $T$, but using gradient flow as smoothing method.}
\label{fig:double_extr_tcorr}
\end{figure}

We first determine the rate following the strategy pursued in past works, namely, to determine the rate from the inversion of the double-extrapolated correlator. This is the procedure we followed:

\begin{enumerate}
\item Numerically-computed lattice correlators $G_L(tT, N_t, n_\cool)/T^5$ are interpolated in $n_\cool$ and $tT$. The first interpolation is necessary to take the continuum limit at fixed smoothing radius, i.e., at fixed $n_\cool/N_t^2$. The second interpolation is necessary to define the correlators obtained on coarser lattices for the same time separations obtainable on the finest one.
\item Interpolated correlators are then extrapolated towards the continuum limit at fixed $n_\cool/N_t^2$ assuming $O(a^2)=O(1/N_t^2)$ leading corrections. Examples are shown in Fig.~\ref{fig:double_extr_tcorr} (top panels).
\item Continuum-extrapolated correlators are finally extrapolated towards the zero-cooling limit $n_\cool/N_t^2 \to 0$ assuming linear corrections in $n_\cool/N_t^2$~\cite{Altenkort:2020axj,Bonanno:2022dru,Bonanno:2022hmz}. Examples are shown in Fig.~\ref{fig:double_extr_tcorr} (bottom left panel).
\item Once the double-extrapolated correlator is obtained, we use the HLT method described in Sec.~\ref{sec:HLT} to invert it and compute $\Gamma_\sphal$. The inversion is performed for various values of the smearing width.
\end{enumerate}

We now comment the obtained results. The final double-extrapolated correlator is shown in Fig.~\ref{fig:double_extr_tcorr} (bottom right panel), where it is compared with the one obtained at the same temperature in~\cite{Kotov:2018aaa}, where the gradient flow was employed as smoothing method.

\begin{figure}[!htb]
\includegraphics[scale=0.333]{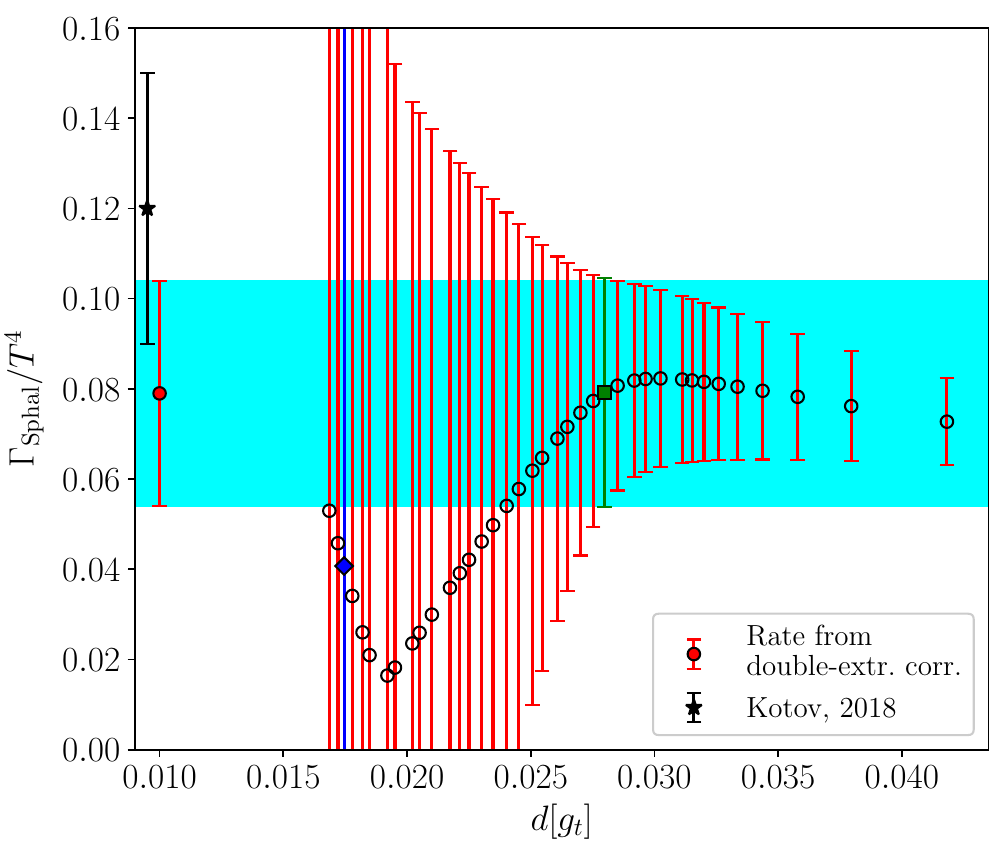}
\includegraphics[scale=0.333]{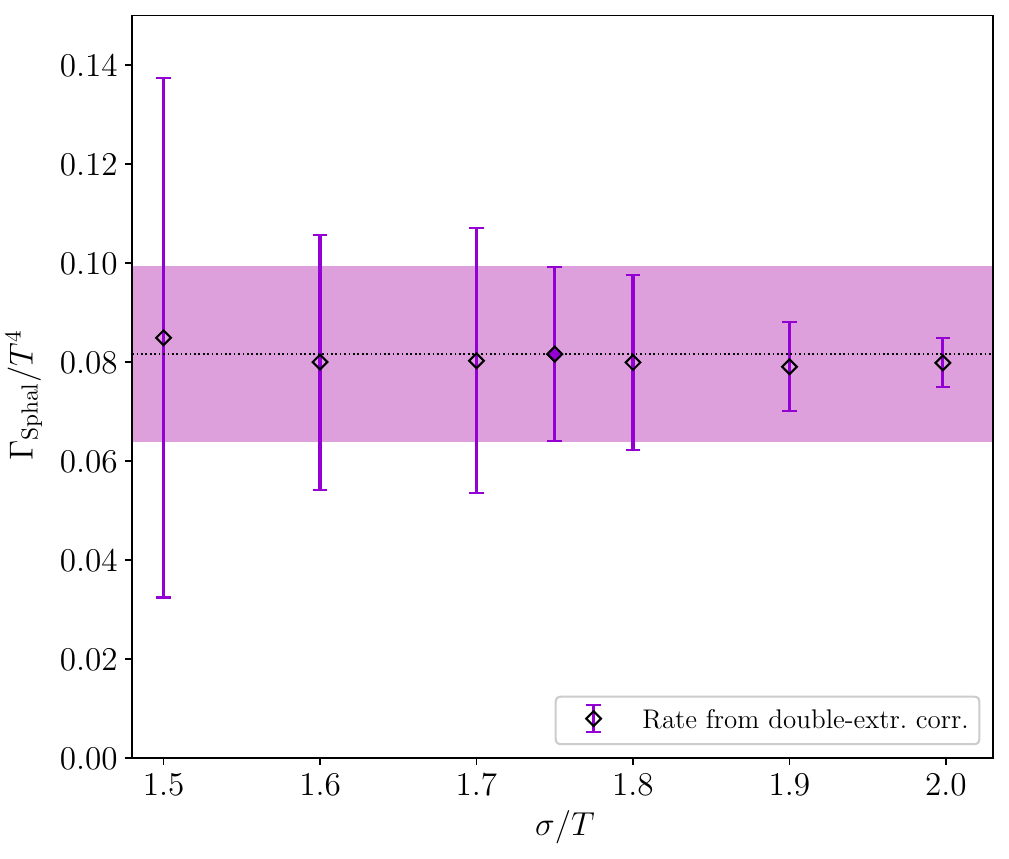}
\caption{Top panel: sphaleron rate from the inversion of the double-extrapolated correlator using the HLT method as a function of the regulator parameter. The horizontal axis is expressed in terms of $d[g_t](\lambda) = \sqrt{A[g_t]/B[g_t]}(\lambda)$. Bottom panel: dependence of the reconstructed sphaleron rate as a function of the smearing width of the target function.}
\label{fig:rate_double_extr}
\end{figure}

In Fig.~\ref{fig:rate_double_extr} we show the results for $\Gamma_\sphal$ as a function of the regulator parameter $\lambda$, which are more conveniently expressed in terms of $d[g_t](\lambda) = \sqrt{A[g_t]/B[g_t]}(\lambda)$. Our reconstruction is more noisy for smaller values of $\lambda$, as expected, and the signal improves for larger values of $\lambda$. In the end, we opt for a conservative estimate of the error, which is depicted as a round point and a shaded area in Fig.~\ref{fig:rate_double_extr}. Our final result is:
\beq\label{eq:rate1f}
\frac{\Gamma_\sphal}{T^4} = 0.079(25), \quad T=1.24 T_c.
\eeq
which is $\sim33\%$ smaller but in agreement with the result of~\cite{Kotov:2018aaa} for the same temperature, 0.12(3). This result was obtained with a smearing width $\sigma/T=1.75$ for the target function, but no significant variation was observed varying this value between 1.5 and 2, cf.~Fig.~\ref{fig:rate_double_extr} (bottom panel).

\FloatBarrier

\FloatBarrier

\subsection{Double extrapolation of the rate}

\begin{figure}[!t]
\includegraphics[scale=0.41]{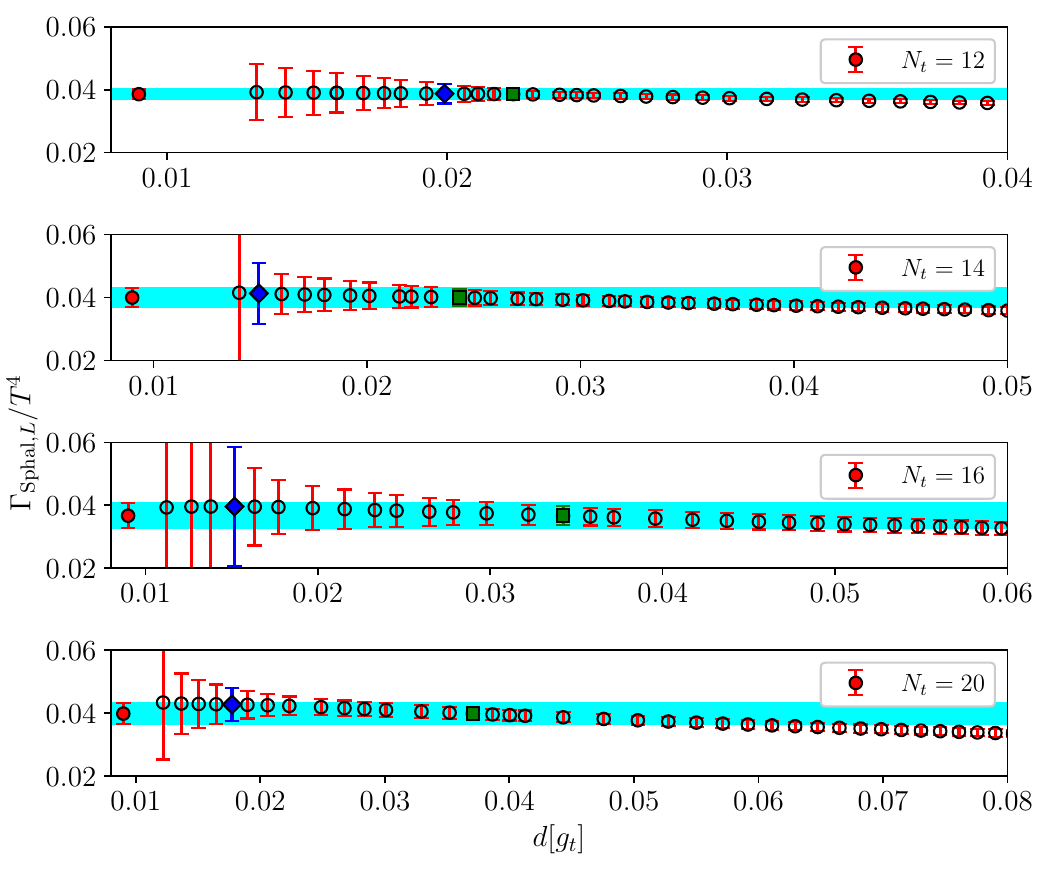}
\includegraphics[scale=0.41]{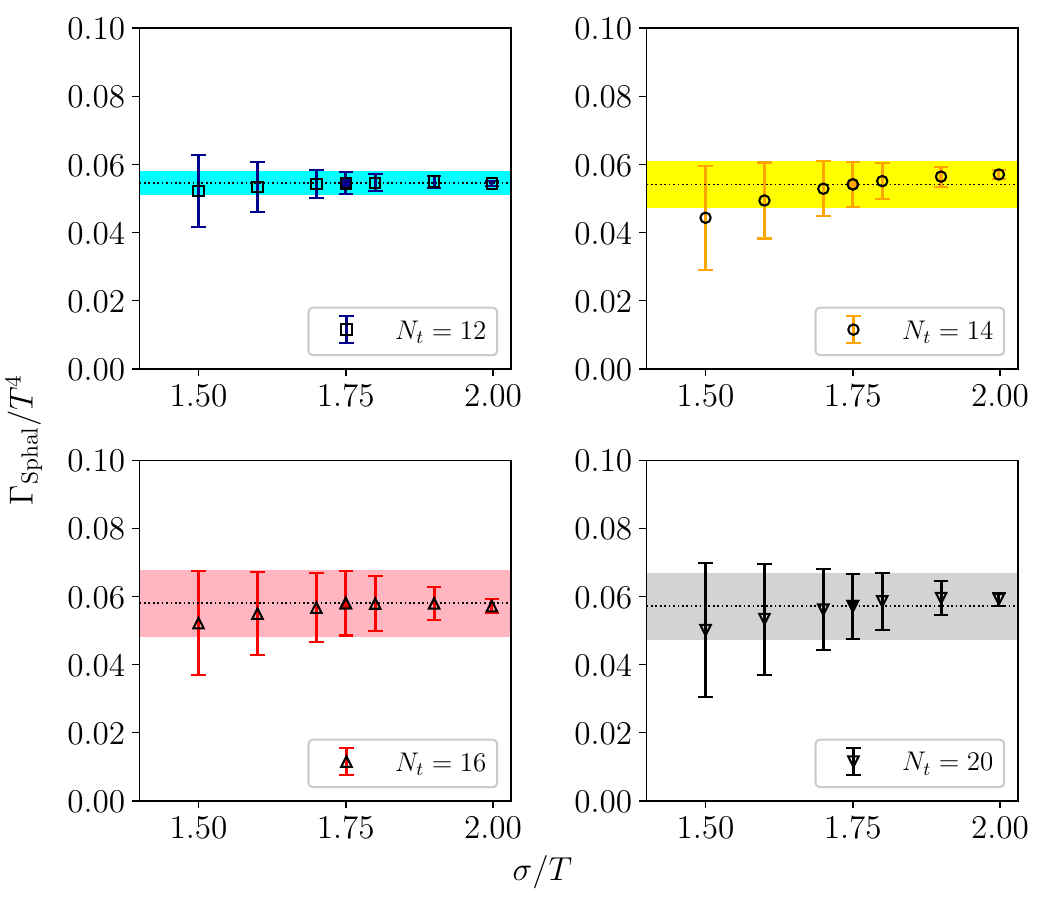}
\includegraphics[scale=0.41]{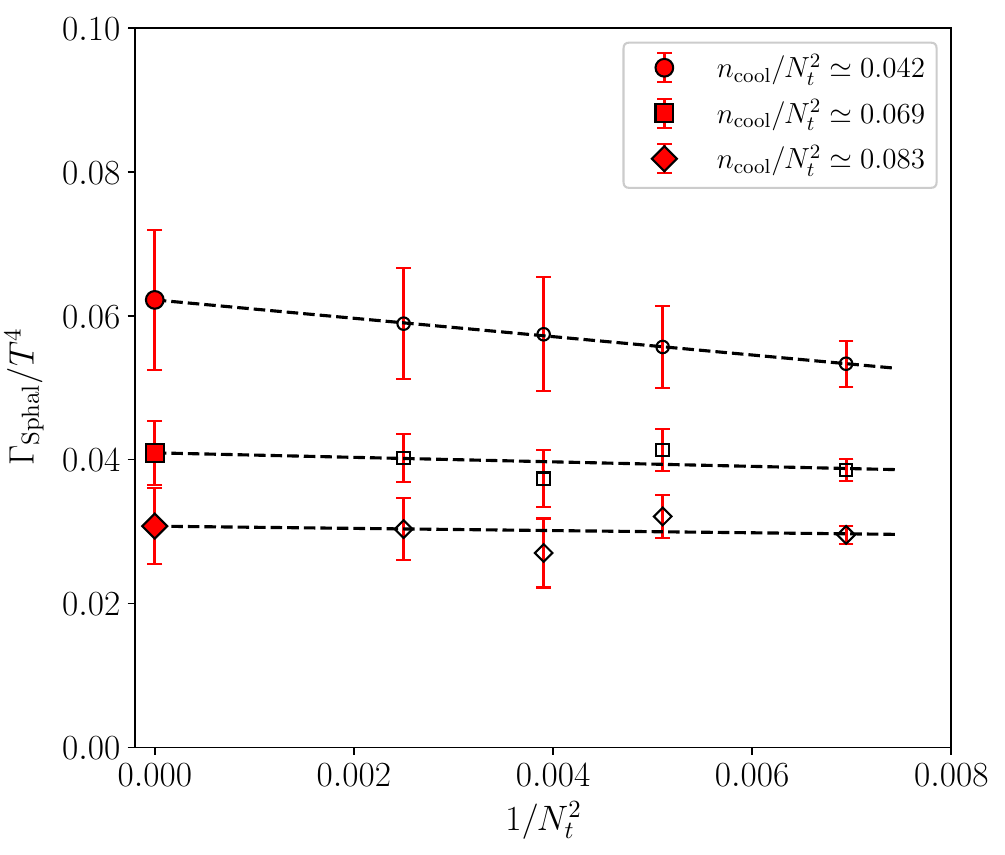}
\caption{Top panel: results for the sphaleron rage $\Gamma_{\sphal,L}$ obtained using the HLT method to invert finite-lattice-spacing and finite-smoothing-radius lattice correlators. Cental panel: $\sigma$-dependence of $\Gamma_{\sphal,L}$ for all available lattice spacings and for a single smoothing radius, corresponding to $n_\cool/N_t^2\simeq 0.04$. Bottom panel: extrapolation towards the continuum limit at fixed smoothing radius of $\Gamma_{\sphal,L}$.}
\label{fig:rate_finite}
\end{figure}

We now move to present results obtained with the alternative strategy proposed in~\cite{Bonanno:2023ljc}:

\begin{enumerate}
\item We extract the sphaleron rate $\Gamma_{\sphal,L}(N_t, n_\cool)$ using the HLT method from the inversion of finite-lattice-spacing and finite-smoothing radius. Obtained results as a function of the regulator parameter $\lambda$, and plotted against $d[g_t](\lambda) = \sqrt{A[g_t]/B[g_t]}(\lambda)$, are shown in Fig.~\ref{fig:rate_finite} (top panel). This approach has two advantages: first, it avoids the necessity of interpolating the correlators in time, second, the result of the inversion is less noisy.
\item Also in this case we adopted $\sigma/T=1.75$ for the smearing width of the target function, and no significant difference was observed varying $\sigma/T$ between 1.5 and 2, cf.~Fig.~\ref{fig:rate_finite} (central panel).
\item Results for $\Gamma_{\sphal,L}(N_t, n_\cool)$ are extrapolated towards the continuum limit at fixed $n_\cool/N_t^2$ assuming $O(a^2)=O(1/N_t^2)$ leading corrections. Examples are shown in Fig.~\ref{fig:rate_finite} (bottom panel). As it can be observed, another advantage of this strategy is that $\Gamma_{\sphal,L}$ is affected by much smaller lattice artifacts compared to the correlator $G_L$.\\
Concerning the smothing radius, in order to fix $n_\cool/N_t^2$ we interpolated $\Gamma_{\sphal,L}$ as a function of $n_\cool$ for each value of $N_t$. However, in this case, we found that also this interpolation is not strictly speaking necessary. As a matter of fact, we found no difference in the obtained results if only integer values of the number of cooling steps are used. In the latter case, this means that, given a certain number of cooling steps $n_\cool$ for a lattice with temporal extent $N_t$, one determines the number of cooling steps $n_\cool'$ for a lattice with temporal extent $N_t'$, corresponding to the same smoothing radius, as $n_\cool' = \mathrm{round}[ n_\cool(N_t'/N_t)^2]$, where $\mathrm{round}[x]$ denotes the rounding to the closest integer to $x$.
\end{enumerate}

\begin{figure}[!t]
\includegraphics[scale=0.45]{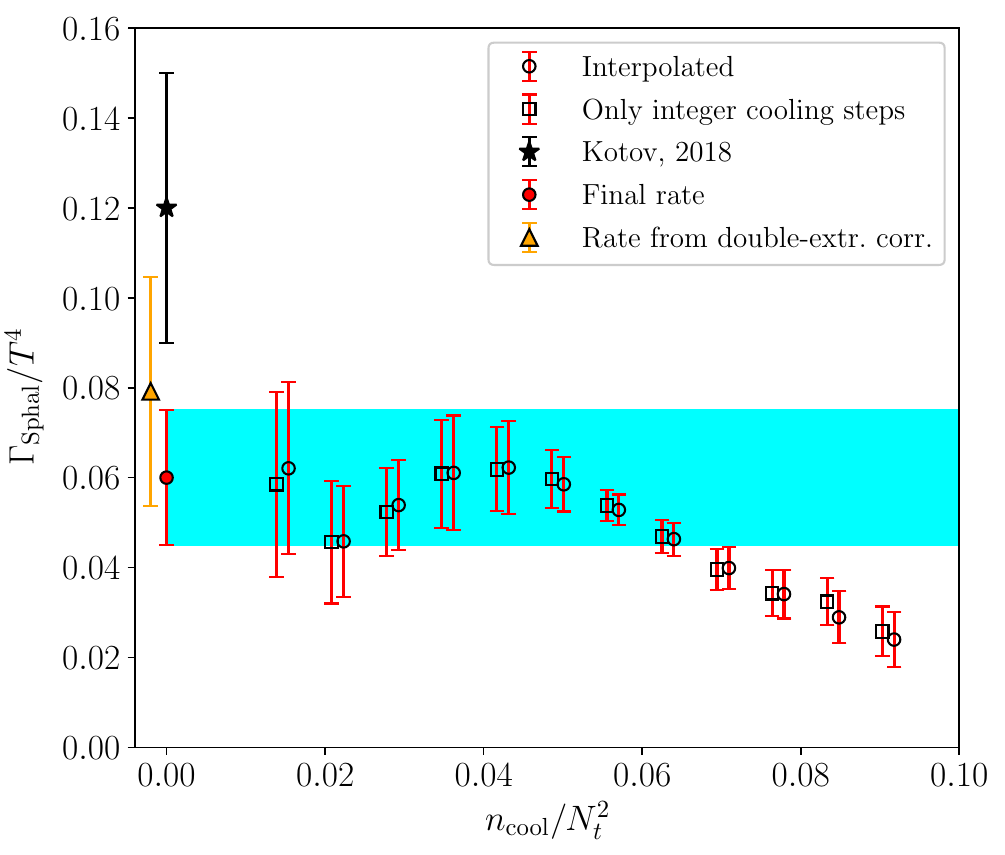}
\caption{Dependence on the smoothing radius of the continuum-extrapolated results of the sphaleron rate obtained by fixing the smoothing radius with and withour interpolation in $n_\cool$ (round and square points respectively). Our final result is represented as a full round point in $n_\cool/N_t^2=0$ and as a shaded area. For the sake of comparison we also report the result obtained from the inversion of the double-extrapolated correlator we obtained in Sec.~\ref{sec:res1} (triangle point), and the result obtained in~\cite{Kotov:2018aaa} (starred point).}
\label{fig:rate_final_res}
\end{figure}

The continuum extrapolations of the sphaleron rate, obtained from the procedure outlined above, are shown as a function of $n_\cool/N_t^2$ in Fig.~\ref{fig:rate_final_res}. As expected, for sufficiently small values of $n_\cool/N_t^2$, we observed that the sphaleron rate does not show a significant dependence on the smoothing radius. This is reasonable, as the smoothing radius is the zero-frequency limit of the slope of the spectral density of $G(t)$, i.e., it is expected to be dominated by the long-distance tails of the correlator, and to be highly insensitivity to the short-distance behavior of $G(t)$, which is the mostly influenced by smoothing.

Thus, a plateau in the sphaleron rate as a function of $n_\cool/N_t^2$ signals an effective separation between these two scales, i.e., the UV scale introduced by cooling, and the IR scale of topological fluctuations which contribute to $\Gamma_\sphal$. This behavior closely resembles that of the topological susceptibility as a function of the gradient flow time.

In conclusion, in this case we do not perform any zero-cooling extrapolation of $\Gamma_\sphal$, and as our final result for the sphaleron rate we simply take the value at the plateau, cf.~Fig.~\ref{fig:rate_final_res}. Keeping into account the variations of the central values of the points in the plateau, and their statistical errors, we give the following final estimate:
\beq\label{eq:rate2f}
\frac{\Gamma_\sphal}{T^4} = 0.060(15), \quad T=1.24 T_c.
\eeq

This result is in perfect agreement with the one obtained from the inversion of the double-extrapolated correlator, cf.~Eq.~\eqref{eq:rate1f}, but it is more accurate. Also, this results confirms what was found in Sec.~\eqref{sec:res1}, i.e., a smaller result for $\Gamma_\sphal$ compared to the one found in~\cite{Kotov:2018aaa} for the same temperature $T\simeq 1.24 T_c$ explored here.

On the other hand, our final result $\Gamma_\sphal=0.060(15)$ is in excellent agreement with the result obtained in~\cite{BarrosoMancha:2022mbj} for a close-by temperature, $T\simeq 1.30 T_c$, $\Gamma_\sphal=0.061(2)$. This result was obtained following a completely different strategy (based on the computation of the susceptibility of the ``sphaleron topological charge'').

\section{Conclusions}\label{sec:conclu}

The problem of determining the sphaleron rate from Euclidean lattice correlators of the topological charge can be formulated as an inverse problem. In this paper we have discussed the main results of~\cite{Bonanno:2023ljc}, where this inverse problem was numerically solved using the recently-introduced HLT method from the Rome group.

We presented two calculations: one consisting of computing $\Gamma_\sphal$ from the inversion of the double-extrapolated correlator of the topological charge density, the other consisting of inverting finite-lattice-spacing and finite-smoothing radius correlators, postponing the double-extrapolation. In both cases we are able to control several sources of systematical errors (finite lattice spacing, finite smoothing radius, dependence on the regulator parameters of the HLT method), and both methods give perfectly compatible results for $\Gamma_\sphal$.

Our final number for the sphaleron rate turns out to be smaller (although compatible within less than two sigmas) with the result of~\cite{Kotov:2018aaa} for the same temperature explored here $T=1.24T_c$. On the ohter hand, we found a perfect agreement with the recent determination of~\cite{BarrosoMancha:2022mbj} for a very close temperature $T=1.3T_c$ (based on the computation of the susceptibility of the ``sphaleron topological charge''). We stress that the small tension with~\cite{Kotov:2018aaa} is not due to the different smoothing methods adopted (cooling here, gradient flow in~\cite{Kotov:2018aaa}), as we find perfectly agreeing results for the double-extrapolated correlator.

Concerning the comparison between the two calculations here presented, we found the second method (based on the inversion of finite-$a$ and finite-$r_s$ correlators) to present several advantages compared to the first one (based on the inversion of the double-extrapolated correlator). More precisely, we found the second strategy to be affected by much smaller lattice artifacts and by a milder dependence of continuum-extrapolated results on the choice of the smoothing radius, and to yield an easier reconstruction of the spectral density from the HLT method. For this reason, the second method turns out to be more feasible for applications in the more computationally-demanding case of full QCD. As a matter of fact, applying the latter method we were able to obtain the first full QCD results for the sphaleron rate, which can be found in~\cite{Bonanno:2023thi}.

\section*{Acknowledgements}
The work of Claudio Bonanno is supported by the Spanish Research Agency (Agencia Estatal de Investigación) through the grant IFT Centro de Excelencia Severo Ochoa CEX2020-001007-S and, partially, by grant PID2021-127526NB-I00, both funded by MCIN/AEI/10.13039/501100011033. Claudio Bonanno also acknowledges support from the project H2020-MSCAITN-2018-813942 (EuroPLEx) and the EU Horizon 2020 research and innovation programme, STRONG-2020 project, under grant agreement No 824093. Numerical simulations have been performed on the \texttt{MARCONI} and \texttt{Marconi100} machines at CINECA, based on the agreement between INFN and CINECA, under projects INF22\_npqcd and INF23\_npqcd.

\end{document}